\documentclass{aa}  

\usepackage{graphicx}
\usepackage{float}
\usepackage{txfonts}
\usepackage[colorlinks,citecolor=blue]{hyperref}
\usepackage{ulem}
\usepackage{subcaption}

\newcommand{\rlc}{R_{\rm LC}}
\usepackage{color}

\begin{document}

   \title{Intra-pulse variability induced by plasmoid formation in pulsar magnetospheres}

   \titlerunning{Intra-pulse variability induced by plasmoid formation in pulsar magnetospheres}
   
   \author{I. Ceyhun Andaç          \inst{1,2}
          \and
          Beno\^{i}t Cerutti\inst{1}
          \and
          Guillaume Dubus\inst{1}
          \and
          K. Yavuz Ekşi\inst{2}
          }

    \institute{Univ. Grenoble Alpes, CNRS, IPAG, 38000 Grenoble, France\\
    \email{benoit.cerutti@univ-grenoble-alpes.fr}
    \and
    Istanbul Technical University, Faculty  of Science  and  Letters,  Physics Engineering  Department, 34469,  Istanbul, Turkey
    }

   \date{Received ???; accepted ???}

 
  \abstract
   {Pulsars show irregularities in their pulsed radio emission that originate from propagation effects and the intrinsic activity of the source.}
   {In this work, we investigate the role played by magnetic reconnection and the formation of plasmoids in the pulsar wind current sheet as a possible source of intrinsic pulse-to-pulse variability in the incoherent, high-energy emission pattern.}
   {We used a two-dimensional particle-in-cell simulation of an orthogonal pulsar magnetosphere restricted to the plane perpendicular to the star spin axis. We evolved the solution for several tens of pulsar periods to gather a statistically significant sample of synthetic pulse profiles.}
   {The formation of plasmoids leads to strong pulse-to-pulse variability in the form of multiple short, bright subpulses, which appear only on the leading edge of each main pulse. These secondary peaks of emission are dominated by the dozen plasmoids that can grow up to macroscopic scales. They emerge from the high end of the hierarchical merging process occurring along the wind current layer. The flux of the subpulses is correlated with their width in phase. Although the full-scale separation is not realistic, we argue that the simulation correctly captures the demographics and the properties of the largest plasmoids, and therefore of the brightest subpulses.}
   {The prediction of subpulses at specific pulse phases provides a new observational test of the magnetic reconnection scenario as the origin of the pulsed incoherent emission. High-time-resolution observations of the Crab pulsar in the optical range may be the most promising source to target for this purpose.}


   \maketitle
%

\section{Introduction} \label{sec:intro}

Rotation-powered pulsars are observed in a wide range of the electromagnetic spectrum from radio waves up to gamma rays \citep{2022ASSL..465...57H}. They exhibit significant fluctuations in their pulse profiles as well as substantial variations in flux \citep{ric69,sie82}. Intra-pulse variability has been reported in the radio band only \citep{ran70,lyn71} because of a lack of sensitivity at other wavelengths. The pulse-to-pulse variability in flux is due to both propagation effects and the intrinsic activity of the source of emission itself. While propagation effects are well known and characterized \citep{goc75, ric84, bla85, goo87, nar92, kum+20}, the origin of the magnetospheric flux variability is still poorly understood. However, it may be tightly connected to the pulsar emission mechanism \citep{mel95}, and therefore the pulse-to-pulse variability carries precious information on how pulsars work and understanding its origin is of prime importance.

Pulsars are well-modeled as rotating, conducting spheres surrounded by a magnetized plasma \citep{gol69}. An electron-positron plasma is created near the neutron star surface and along the open field lines by magnetic conversion \citep{1971ApJ...164..529S}, and possibly in the outer magnetospheric regions by photon-photon annihilation \citep{1996A&A...311..172L}. A large-scale current sheet forms where both magnetic polarities meet beyond the light-cylinder radius \citep{cor90, bog99}, $\rlc$, where the corotation velocity is equal to the speed of light. A fraction of the pulsar spindown is dissipated in the current layer via magnetic reconnection \citep{phi+14a, phi+15a, 2014ApJ...795L..22C, 2015MNRAS.449.2759B, cer+15, cer+17a,2020A&A...642A.204C, 2021arXiv210903935H}. Being long (light-cylinder scale) and thin (plasma skin-depth scale), this layer rapidly breaks up into a chain of plasmoids \citep{zel79, uzd+10}. This fragmentation leads to fast magnetic reconnection and efficient particle acceleration \citep{kag+15}. This may lead in return to intense nonthermal radiation and explain the high-energy pulsed emission \citep{1996A&A...311..172L, 2012MNRAS.424.2023P, 2013A&A...550A.101A, cer+16b, phi+18,kal+18}.

\begin{figure*}
\centering
\includegraphics[width=0.9\textwidth]{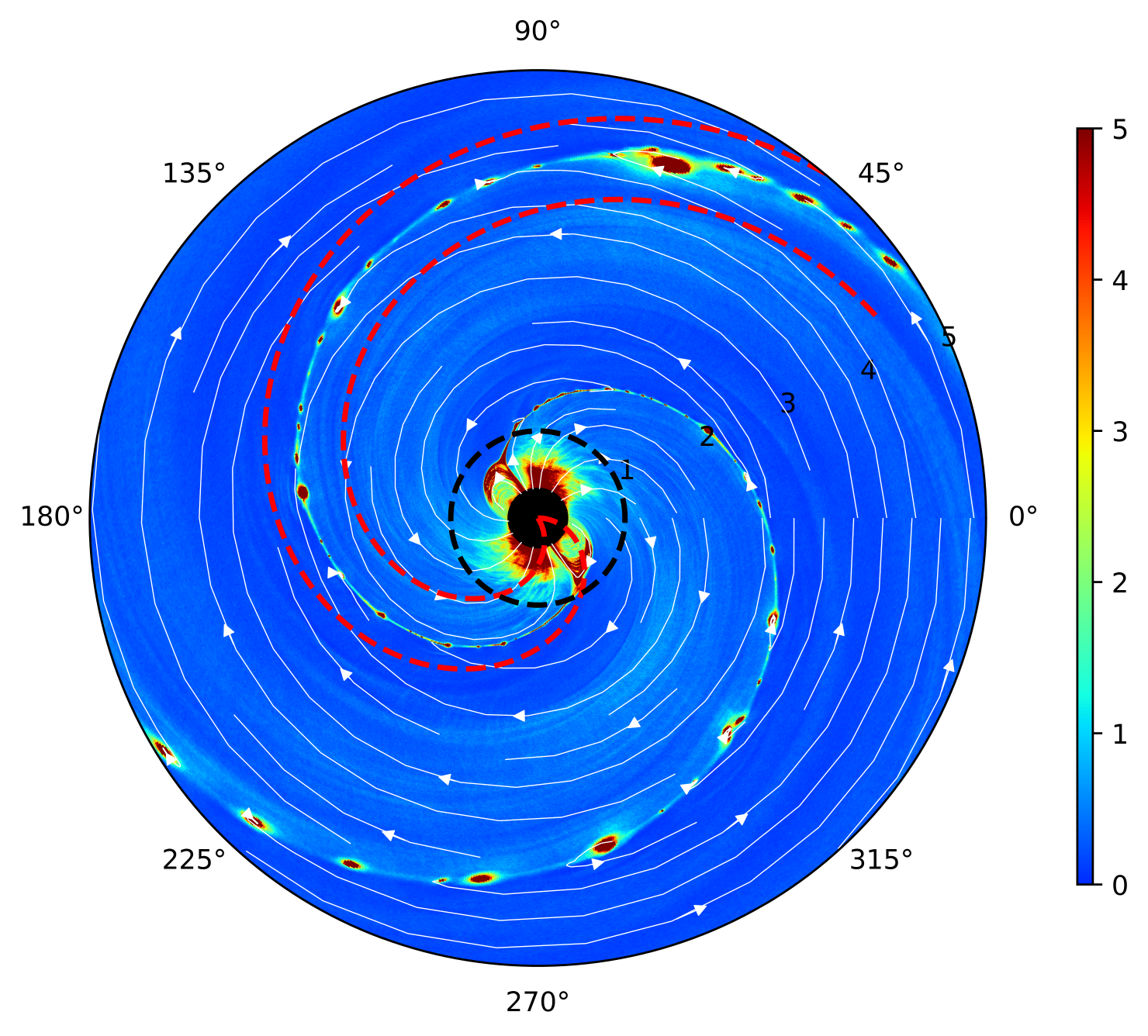}
\includegraphics[width=0.9\textwidth]{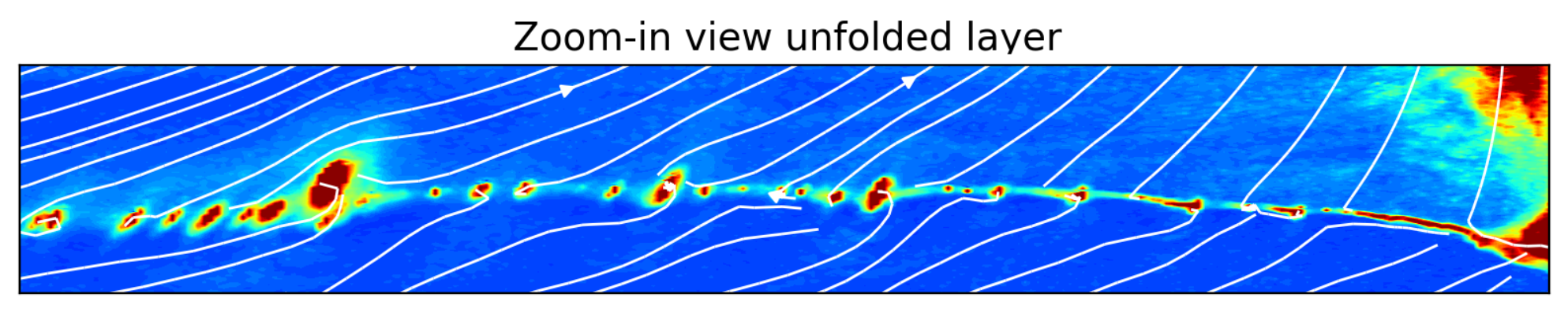}
\caption{Plasma density map ($r$-$\phi$ plane) normalized to the fiducial injected plasma density, $n_0$. Top panel: The black disk is the neutron star and the dashed circle shows the light cylinder located at $\rlc=3r_\star$. Radii are normalized to $\rlc$. The white solid lines and arrows show the magnetic field lines. Bottom panel: Zoomed-in view on one current layer delimited by the red dashed line shown in the top panel. The layer has been unfolded to better observe the chain of plasmoids forming along the spiral.}
\label{fig:spiral}
\end{figure*}

In this work, we investigate whether plasmoid formation and their evolution in pulsar magnetospheres could be a possible source of intrinsic noise in the emission pattern. The intermittent nature of magnetic reconnection as well as the expected broad size distribution of plasmoids \citep{uzd+10, sir+16} make this process a good candidate for a significant source of pulse-to-pulse variability in the observed incoherent emission. We perform a two-dimensional (2D) particle-in-cell (PIC) simulation of an orthogonal pulsar magnetosphere restricted to the equatorial plane as first introduced in \citet{cer+17a}. The simulation is evolved over many pulsar rotation periods and with a refined time-resolution to well characterize the statistical properties of plasmoids and their imprint on the light-curve noise. 

The paper is structured as follows: the numerical methods and the simulation setup are introduced in Sect.~\ref{sect_setup}. Our detailed analysis of the simulation with an emphasis on the formation of plasmoids and their demographics is described in Sect.~\ref{sect_plasmoids}, and the pulse-to-pulse variability and its connection to plasmoid evolution is presented in Sect.~\ref{sect_synchrotron}. In Sect.~\ref{sec:discandconc}, we discuss the implications of this work and propose future observational tests to validate the magnetic reconnection scenario as the origin of the incoherent pulsar emission.

\section{The simulation setup} \label{sect_setup}

The simulation is performed with the {\tt Zeltron} PIC code \citep{cer+13, cer+19a}. We use nearly the same numerical setup as the one presented in \citet{cer+17a}, which we briefly outline below for completeness; the main differences being a smaller radial extent of the box size but a longer integration time.

The numerical box is initialized with a split magnetic monopole configuration \citep{mic73b,bog99}. The magnetic moment, $\boldsymbol{\mu}$, is perpendicular to the neutron star spin axis, $\boldsymbol{\mu}\cdot\boldsymbol{\Omega}=0$, where $\boldsymbol{\Omega}$ is the pulsar angular velocity vector. We use spherical coordinates; however, the simulation is restricted in the equatorial plane $(r,\theta=\pi/2,\phi)$. The grid spacing is logarithmically stretched in the radial direction and uniform in the azimuthal direction. The domain inner boundary is set at the stellar surface, $r_{\rm min}=r_{\star}$. The light-cylinder radius is set to $\rlc=c/\Omega = 3 r_\star$, which corresponds to a millisecond pulsar. The outer boundary is located at $r_{\rm max}=30 r_{\star}=10\rlc$. A damping layer beginning at $r=9\rlc$ absorbs outgoing particles and electromagnetic waves to mimic an open boundary.

The simulation begins in a vacuum state around the neutron star. Field lines at the stellar surface are set in solid rotation by enforcing the co-rotation ideal electric field, $\Vec{E}=-\left(\Vec{\Omega}\times\Vec{r}\right)\times\Vec{B}/c$. The plasma particles ---electrons and positrons in our case--- are injected from the surface at the $\Vec{E} \times \Vec{B}$ drift velocity. An equal number of electrons and positrons is injected with a number density $n_0=10 n_{\rm GJ}$, where $n_{\rm GJ}=\Omega B_\star/2\pi e c$ is the fiducial critical density \citep{gol69}, and $B_{\star}$ is the surface magnetic field at the poles. Pair production is neglected in the simulation. Its effects are mimicked by the injection of plasma at the surface, which is sufficient to establish a filled, quasi-force-free magnetosphere. The fiducial magnetization parameter on the stellar surface is $\sigma_{\star} = B^2_{\star}/4\pi n_{\star}m_{\rm e} c^2 = 250$ where $m_{\rm e}$ is the electron rest mass. The radiation-reaction force is activated in the simulation to capture curvature and synchrotron cooling.

The total number of cells is $2048\times 2048$. Two particles are injected per time-step in every cell located at the surface of the star. The simulation time-step is set to half the Courant-Friedrich-Lewy critical step. There are about $2.58 \times 10^4$ time-steps per spin period, $T_{\rm spin}$. To characterize the intrinsic noise due to plasmoid formation in the wind  current sheet, the simulation is evolved up to $t_{\rm max} \simeq 70 T_{\rm spin}$.

\section{Formation and evolution of plasmoids}\label{sect_plasmoids}

\begin{figure}
\centering
\includegraphics[width=0.5\textwidth]{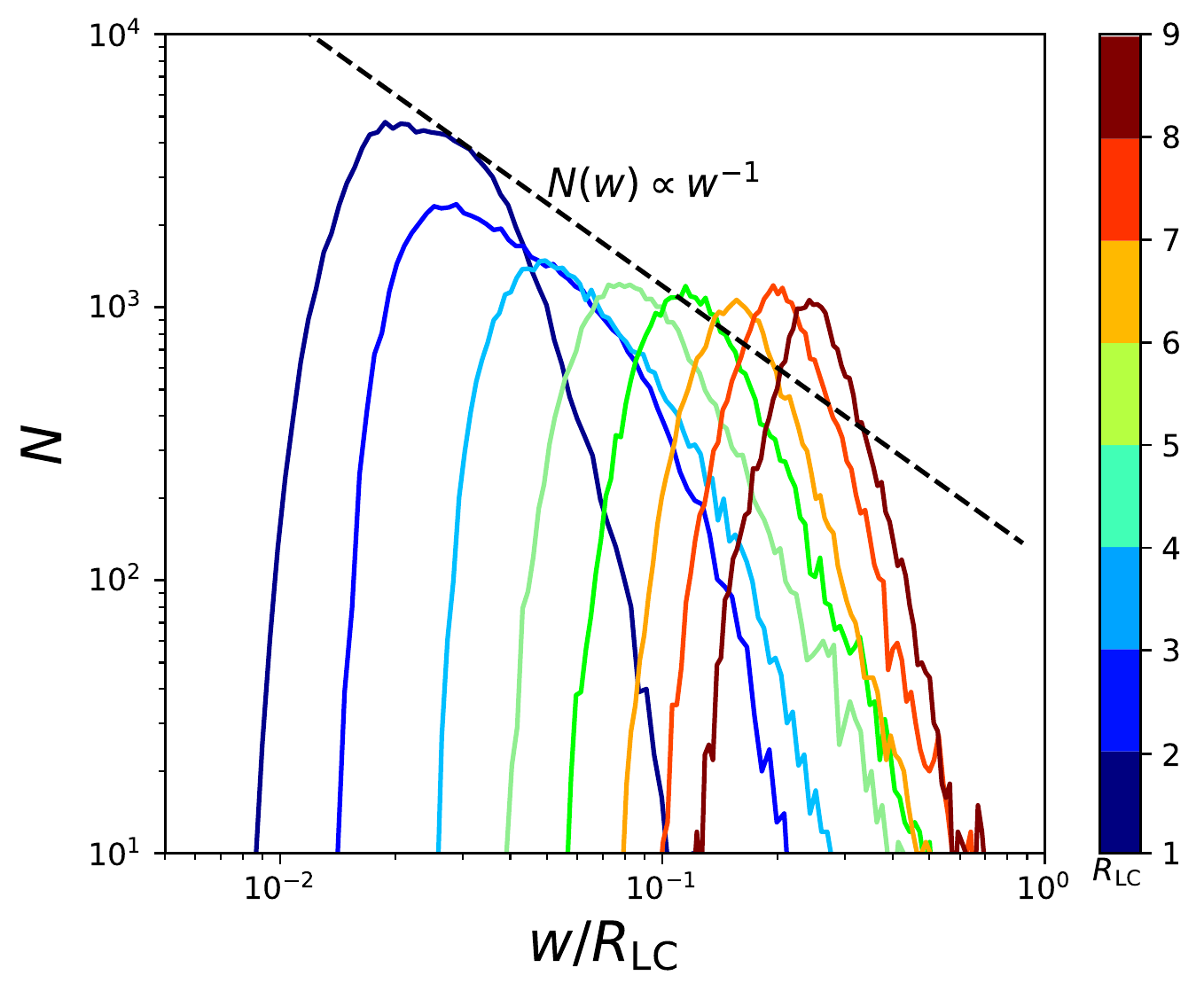}
\caption{Plasmoids size distribution at different radii in units of $\rlc$ (color-coded, ranging from the $1$-$2\rlc$ interval up to $8$-$9\rlc$ interval). The dashed line is $N$ proportional to the inverse of the island width $w$ predicted by \citet{uzd+10}.}
\label{Fig:size-hist}
\end{figure}

After a few pulsar spin periods, the simulation box is filled with plasma and a nearly steady-state force-free magnetosphere is established. The wind current sheet appears as two thin Archimedean spirals separated by $\Delta\phi=\pi$, such that two consecutive arms are equally spaced along the radial direction by $\pi R_{\rm LC}$. Figure~\ref{fig:spiral} shows a representative snapshot of the plasma density at the late stages of the simulation ($t=60 T_{\rm spin}$). The striking feature outlined in \citet{cer+17a} is the fragmentation of the current layers into multiple plasma overdensities trapped in magnetic islands, or plasmoids in the following. 

Nearly all of the plasmoids are created in the vicinity of the light cylinder where the wind current sheet forms and fragments under the action of the tearing instability, mediating fast magnetic reconnection \citep{uzd+10,lou+12}, which in return leads to efficient nonthermal particle acceleration. After their formation, plasmoids dynamically grow by merging with each other in a hierarchical manner along the spiral. Only a few macroscopic structures remain along each spiral at the end of the merging process. For example, in Figure~\ref{fig:spiral}, a giant plasmoid of size on the order of the neutron star itself (full width $w\sim r_{\star}$) lies in the upper spiral at $r\approx 4\rlc$. Given that plasmoids move away from the star at a significant fraction of the speed of light, the chain of plasmoids must reform continuously at the base of the layer near the light cylinder. Therefore, in steady-state, a given radius statistically probes a different phase of the merging process (see the zoomed-in view on the current layer, the bottom panel in Figure~\ref{fig:spiral}). This process is frozen by relativistic expansion of the wind beyond $r\gtrsim 5 \rlc$ \citep{cer+17a, 2021A&A...656A..91C}.

In order to better quantify plasmoid formation and evolution, we performed a systematic measurement of the width of all plasmoids being produced in the simulation. Figure~\ref{Fig:size-hist} shows the size distribution of plasmoids, characterized by their full width $w$ as a function of radius within a constant $1\rlc$ bin size outside the light cylinder. Plasmoids are identified along each spiral arm as local maxima of the plasma density. The plasmoid width is then inferred from the full width at half maximum of the density profile with respect to the center of each plasmoid. Their width is measured in the simulation frame along the azimuthal direction to avoid length contraction along the radial direction due to the relativistic motion of the pulsar wind, which becomes significant beyond a few light-cylinder radii (see, e.g., the oblate shape of plasmoids at large radii in the top panel of Figure~\ref{fig:spiral}). The plasmoid size distribution is reconstructed from the entire dataset, cumulating a total of more than $10^4$ islands. Island sizes range from about $2\times 10^{-2}\rlc$ up to $0.3\rlc\approx r_{\star}$. Close to the light cylinder, the distribution peaks at $w_{\rm min}\approx 2\times 10^{-2}\rlc$. With a layer thickness of $\delta\approx 3\times 10^{-3}\rlc$ at the light cylinder, it is consistent with the most unstable tearing mode $w_{\rm min}\sim\lambda_{\rm TI}=2\pi\sqrt{3}\delta$. The size distribution within $2$-$3\rlc$ is the broadest and shows hints of a $N\propto w^{-1}$ dependence as predicted by \citet{uzd+10} for a hierarchical merging process. The small-scale separation achieved in the simulation does not allow us to probe a sufficiently broad distribution to constrain the power-law index accurately. At larger radii, the distribution becomes narrower (logarithmically speaking) and freezes. The monotonic shift of the distribution with radius (i.e., $w\propto r$) is entirely explained by the expansion of the wind. The number of plasmoids decreases to almost one-quarter of the initial number with respect to the initial distribution, meaning that plasmoids undergo two merger events on average by the time they escape from the simulation box.

\section{Pulse and subpulse variability} \label{sect_synchrotron}

\begin{figure}[h!]
\centering
\includegraphics[width=0.5\textwidth]{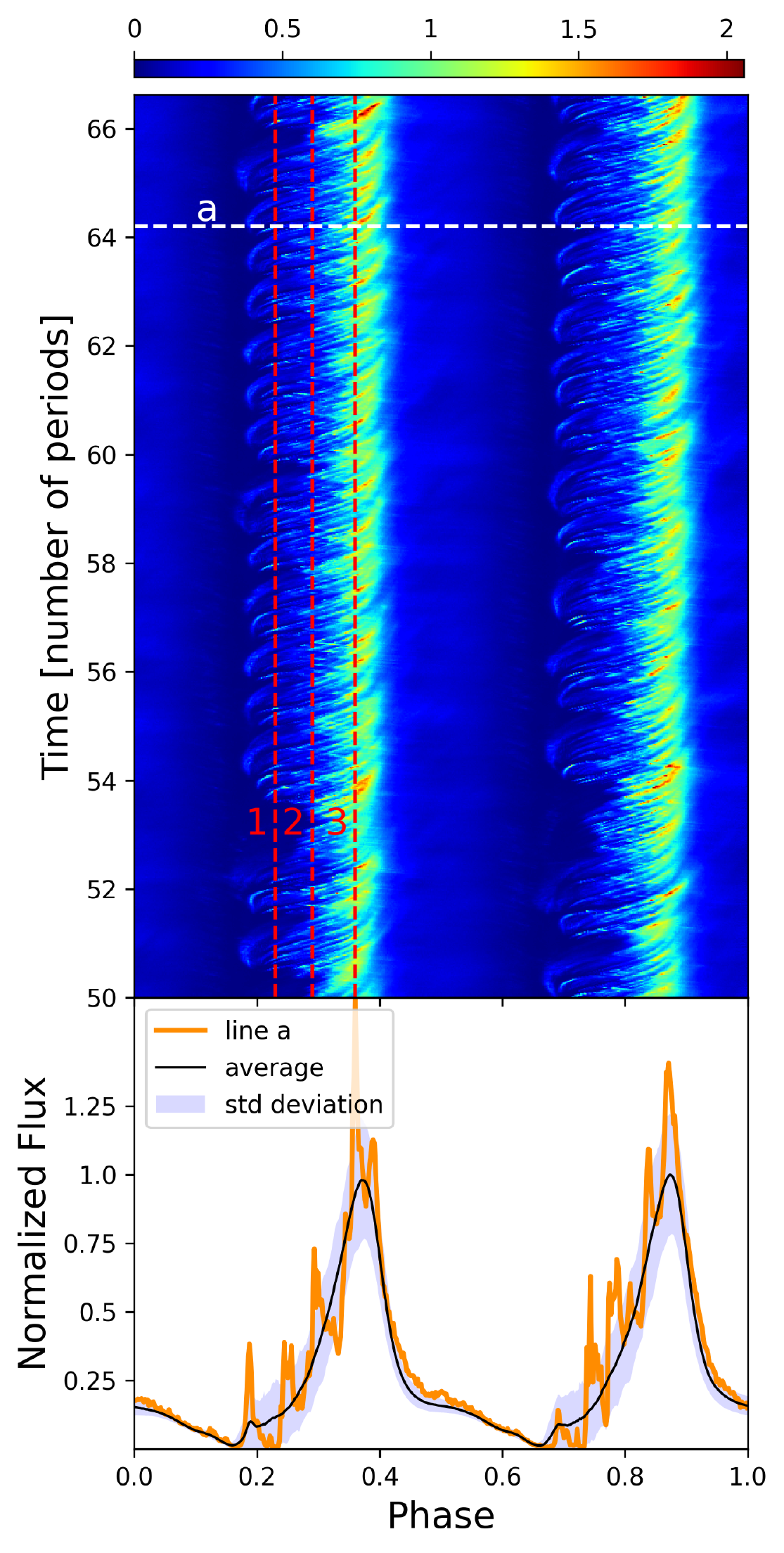}
\caption{Synthetic synchrotron pulse profiles. Top: Phase-time diagram of the synchrotron flux. Vertical red dashed lines numbered 1, 2, and 3 are different phases in the left caustic, $\Phi=0.23$, $0.29$, and $0.36,$ respectively, and are used in the following figures. Bottom: Pulse profile seen by a single observer (orange solid line) generated by crossing the phase-time diagram along the white dashed line labeled ``a'' in the upper panel. The black solid line shows the pulse profile averaged over times $t\geq 50 T_{\rm spin}$. The blue shaded region is bounded by the mean flux which is augmented or diminished by the standard deviation to highlight the variable parts of the pulse profile.}
\label{fig:LC}
\end{figure}

\begin{figure}
\centering
\includegraphics[width=0.5\textwidth]{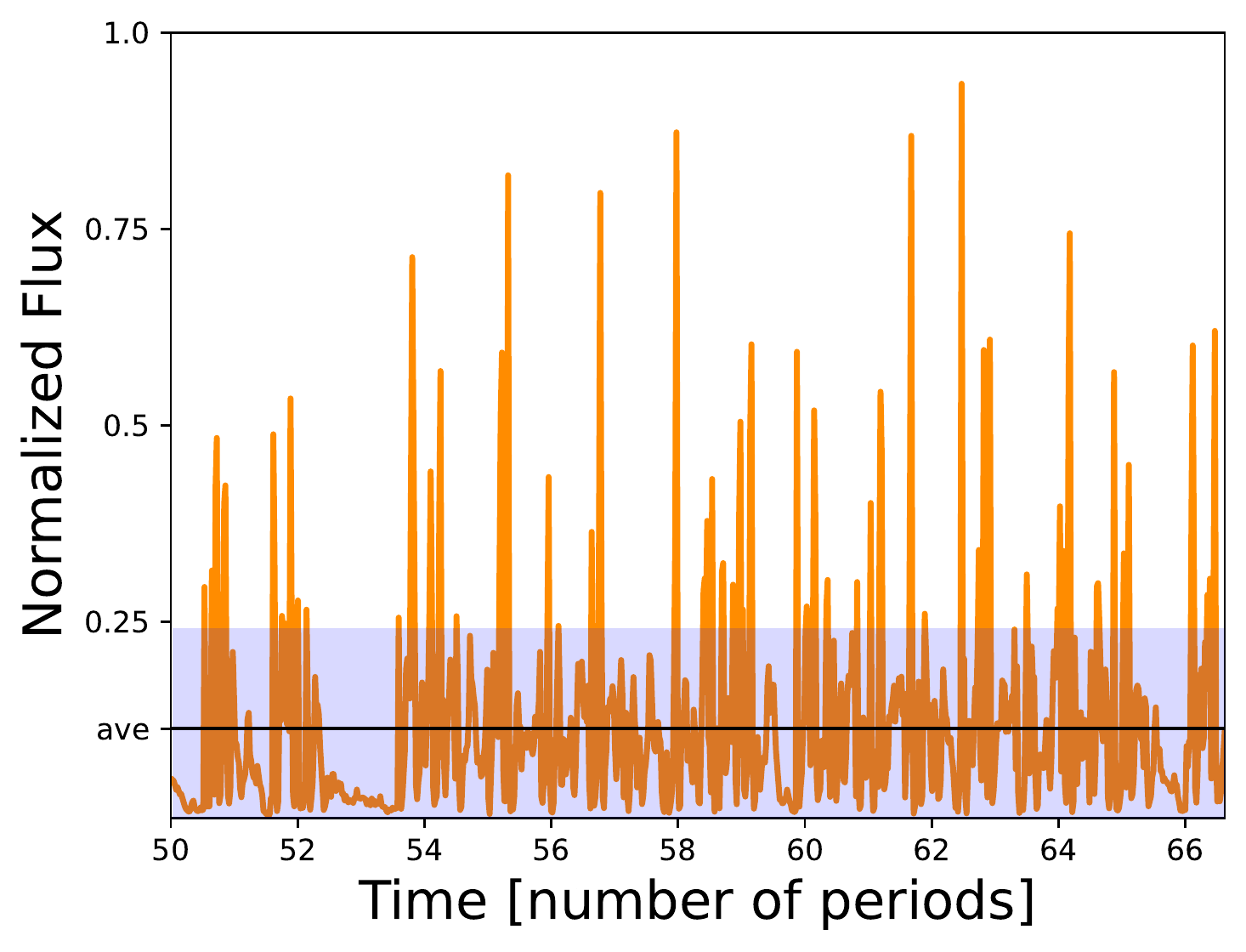}
\caption{Synchrotron flux as a function of time after 50 periods at the pulsar phase $\Phi=0.23$ labeled ``1'' in \autoref{fig:LC}. The average flux is given by the solid black line. The shaded area shows the standard deviation with respect to the mean.}
\label{fig:flux-var}
\end{figure}

\begin{figure}
\centering
\includegraphics[width=0.5\textwidth]{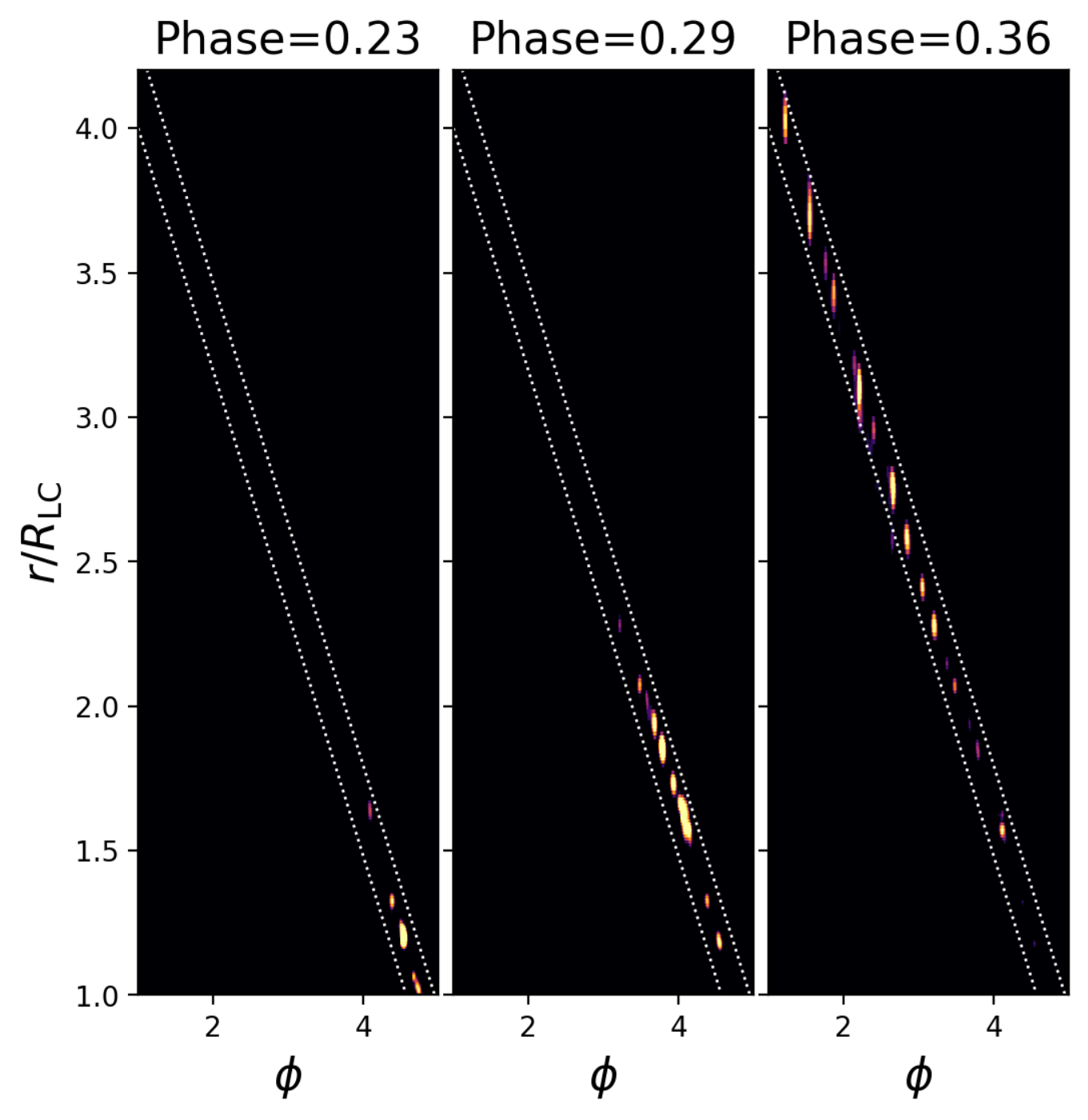}
\includegraphics[width=0.5\textwidth]{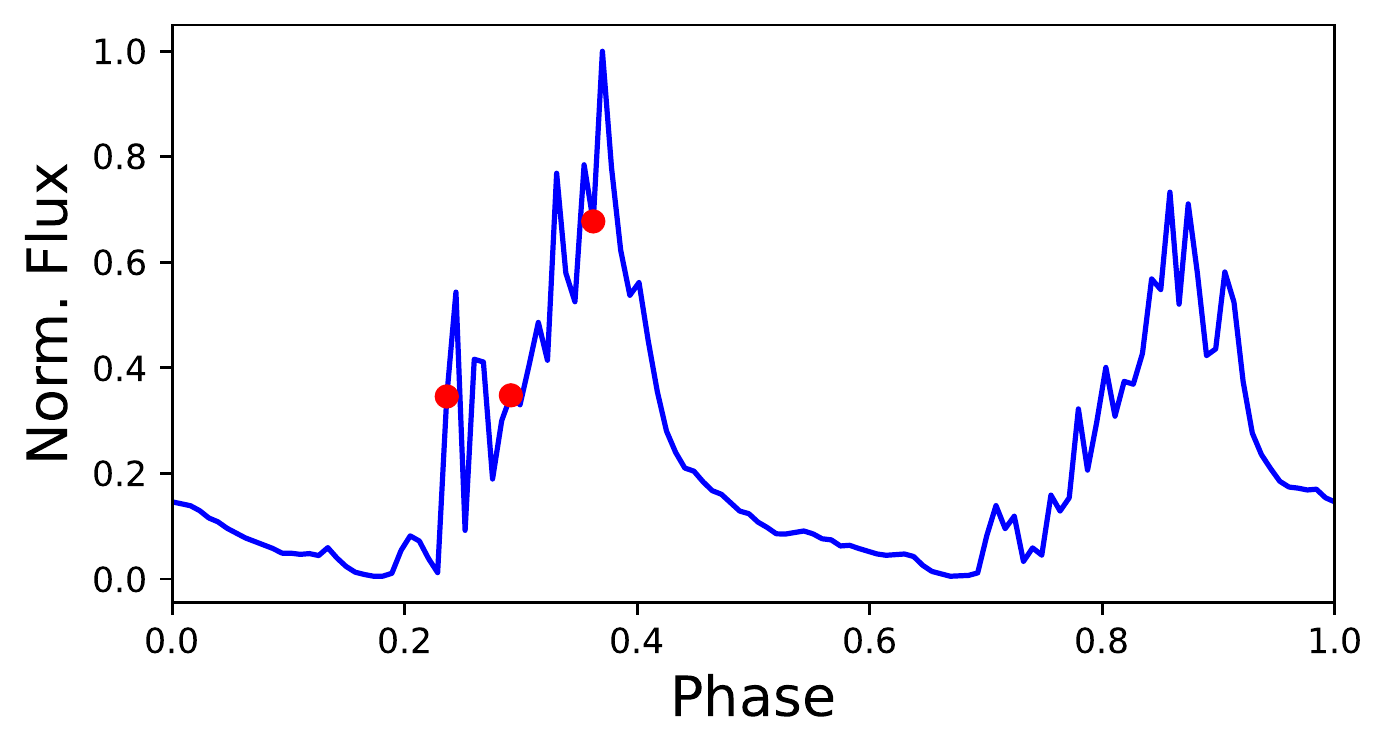}
\caption{Shining plasmoids and origin of subpulses. Top: Map of the synchrotron flux from the current sheet received by the observer at phases 1 (left), 2 (middle), and 3 (right). The oblique white dotted lines are Archimedean spirals encompassing the wind current layer to guide the eye. Bottom: Observed pulse profile. The red dots highlight the phases for which the flux maps are displayed.}
\label{fig:caustic}
\end{figure}

In this section, we evaluate the impact of plasmoid formation on the synchrotron emission pattern. Following \citet{cer+16b}, we compute the synthetic synchrotron pulse profile from the simulation particles and fields as seen by an ensemble of observers located at infinity and along the equatorial plane. We recover the main expected features: two bright, nearly identical and symmetric pulses of light,  each of about $\Delta\Phi\sim 0.2$  width in pulsar phase, and separated by $0.5$ in phase. This pattern stems from the radiation emitted in the current layer. Taking into account the finite photon time of flight, the region of the magnetosphere that is being probed by an observer at a given phase is an Archimedean spiral. Whenever this spiral overlaps with the current sheet spiral (also close to being perfectly Archimedean), there is a pulse of light, which therefore means there are two pulses per period. This is the caustic effect in the wind current layer identified in \citet{cer+16b}. The average pulse profile is shown by the black solid line in Figure~\ref{fig:LC} (bottom panel).

The novel feature highlighted here is the formation of bright, short subpulses on top of the main pulse profile. These structures are intermittent and appear only on the leading edge of each pulse. The orange solid line in Figure~\ref{fig:LC} shows the light curve seen by an observer during one rotation period. It is composed of about ten easily identifiable bright subpulses per pulse, each of about $\delta\Phi\sim 0.01$ width in pulsar phase (i.e., about 10\% of the main pulse leading edge width). A quantitative analysis of the full dataset shows that the average number of major subpulses per pulse is $11\pm 3$. Major subpulses are defined as those with a flux exceeding the mean plus one standard deviation. They appear in Figure~\ref{fig:LC} (top panel) as oblique filaments, meaning that these subpulses are actually regions of enhanced emission that are drifting in phase. Their drifting motion is too fast to be detected by a single, fixed observer. For this reason, subpulses appear randomly from one period to another. Figure~\ref{fig:flux-var} shows the flux as a function of time, measured at the fixed pulsar phase $\Phi=0.23$ (phase ``1'' in Figure~\ref{fig:LC}) where the flux is most variable. At this phase, the standard deviation is similar to the mean flux. It is also a phase at which subpulses seem to form quasi-periodically on a timescale on the order of the pulsar period (see the top panel in Figure~\ref{fig:LC}).

The origin of subpulses is intimately related to plasmoid formation in the pulsar wind. Figure~\ref{fig:caustic} maps the emitting regions at three phases selected in the leading edge of the first pulse. At the base of the pulse (phase 1), the observer sees plasmoids shining in the innermost parts of the current sheet only. This region is very active in terms of the formation and merger of plasmoids, explaining the high level of variability reported above. At phases 2 and 3, more plasmoids located further away begin to contribute to the full flux. At the pulse maximum (phase 3) and beyond in the decaying part of the pulse, all plasmoids contribute within the limits of our box size. Given that the synchrotron flux is strongly peaked near the light cylinder, where the field is strongest, a larger box size would make a negligible difference. The pulse profile shows little variability in the decaying part because the total flux is averaged out by the superposition of emission from all plasmoids along the spiral, and therefore is not very sensitive to large individual plasmoids. This explains the asymmetric nature of the variability pattern in each pulse. At phase 1, the flux can sometimes be below average for an appreciable amount of time; see for example the flux minima at $t\sim 53 T_{\rm spin}$ and $t\sim 66 T_{\rm spin}$ (Figures~\ref{fig:LC}-\ref{fig:flux-var}). This phenomenon is related to an episodic thickening of the current layer near the light cylinder, which temporarily prevents the layer from fragmenting and forming new plasmoids. These quiet episodes are usually followed by a recovery phase where the flux is well above average at and around phase 3 (e.g., $t\sim 54 T_{\rm spin}$), because the layer eventually breaks up at it stretches outwards, leading to a delay in the phase of the emission. This episodic thickening of the current sheet may be a numerical artefact due to the unrealistically small scale-separation achieved in the simulation.

\begin{figure}
\includegraphics[width=0.5\textwidth]{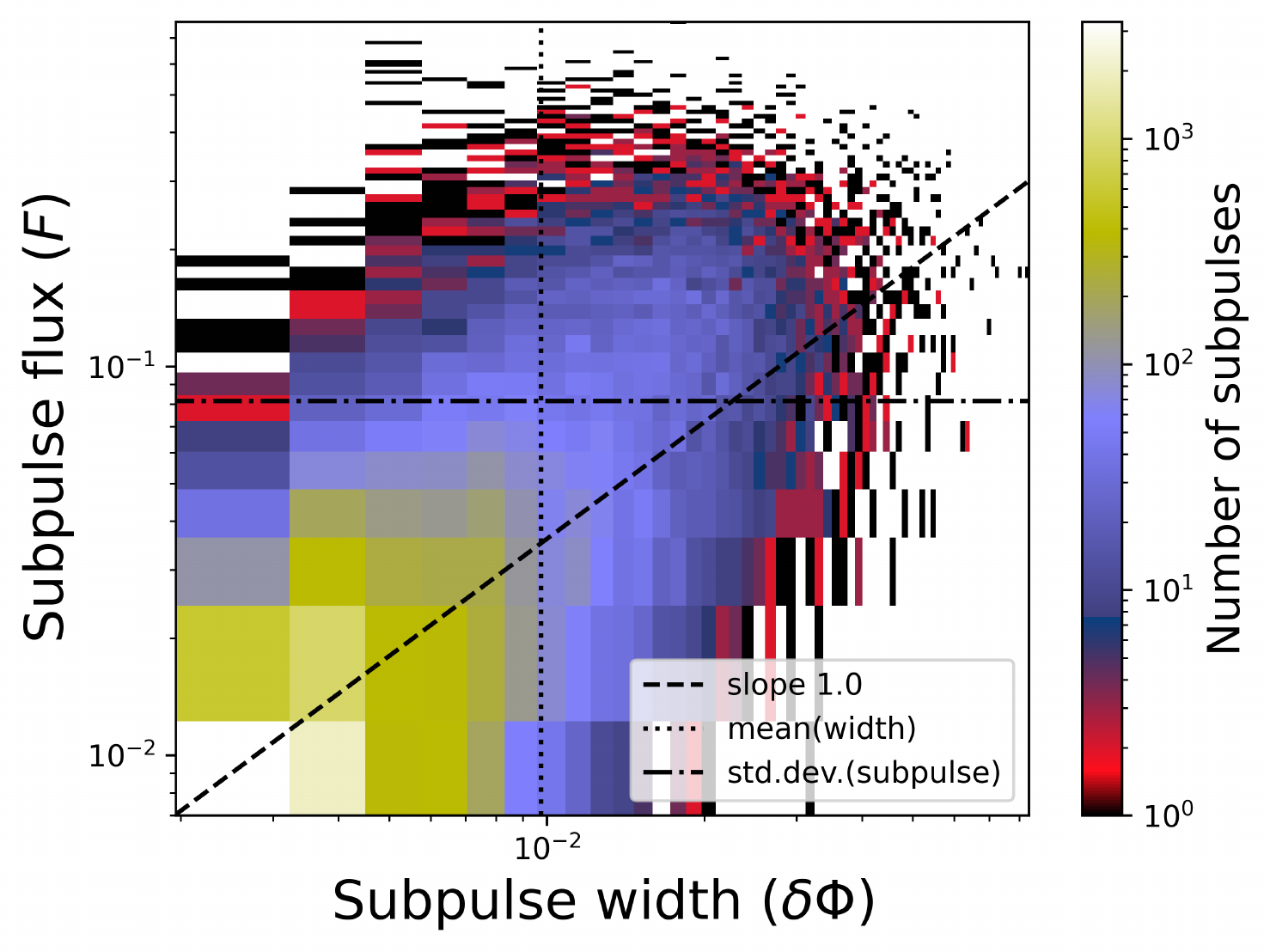}
\includegraphics[width=0.45\textwidth]{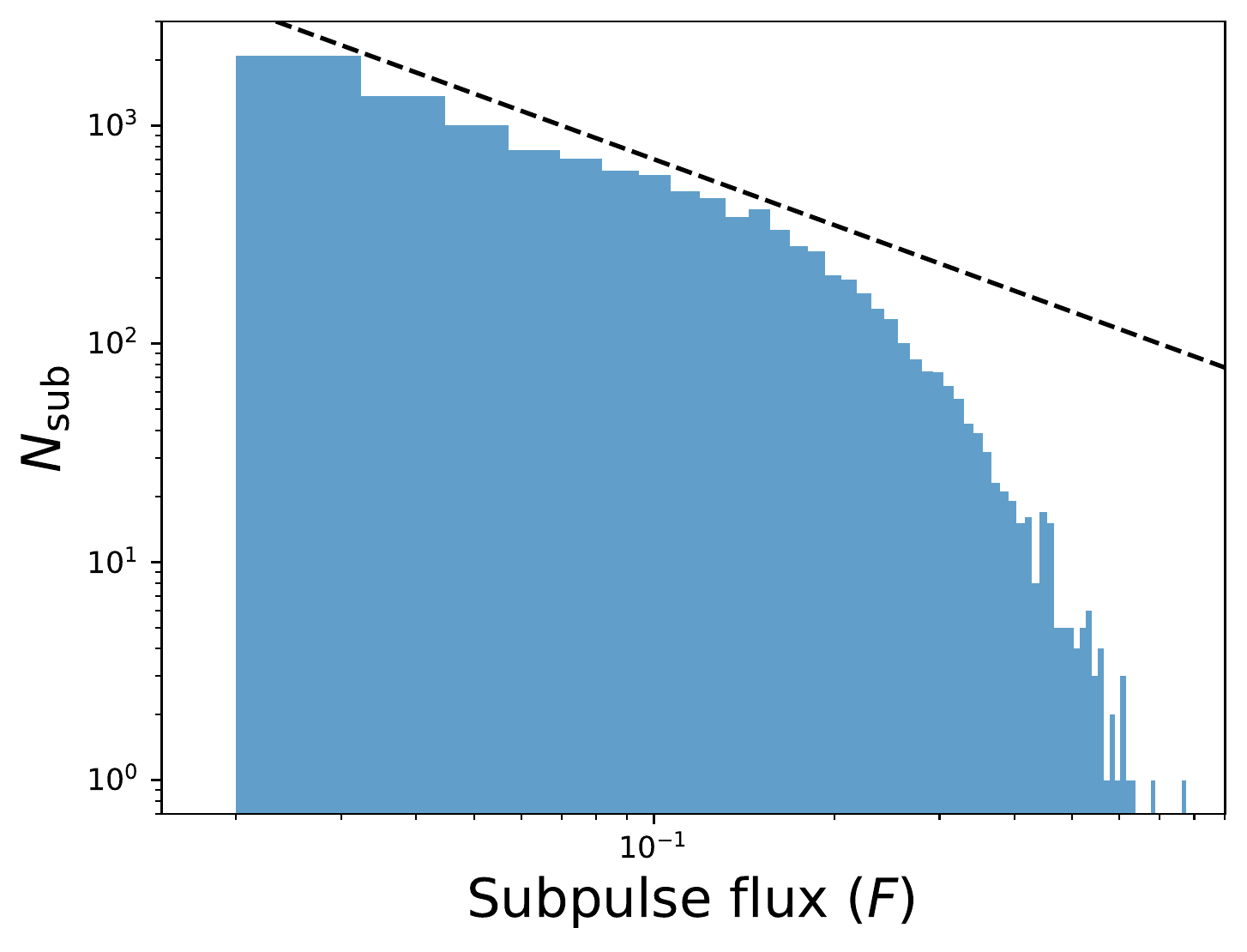}
\caption{Demographics of all synchrotron subpulses in the simulation dataset. Top: Subpulse width in phase ($\delta\Phi$) and maximum flux with respect to the mean profile ($F$) distributions. The dashed oblique line is $F\propto \delta\Phi$. Bottom: Subpulse flux distribution. The black dashed line shows $N_{\rm sub}\propto F^{-1}$.}
\label{fig:subpulse}
\end{figure}

The top panel of Figure~\ref{fig:subpulse} reports on the distribution of maximum flux, $F$, and width in pulsar phase, $\delta\Phi$, for all subpulses measured in the full dataset. The subpulse flux is measured with respect to the mean profile, and is normalized to the maximum of the mean flux. Subpulses can be nearly as prominent as the average pulse profile itself. The distribution is broad, but a clear trend emerges: the brighter the subpulse is, the broader it is in phase. The proportionality relation between these two quantities is consistent with expectations for emission from plasmoids. The total synchrotron power in the subpulse is on the order of $P_{\rm sub}\sim F\times\delta\Phi$. Considering a plasmoid located at a given radius $r$, the synchrotron power it emits will be proportional to the total number of particles it contains, and therefore to its volume. In full 3D, plasmoids become flux ropes, extending at most within the full latitudinal ranges of the stripes. For an orthogonal rotator, as modeled here, the volume of the  flux rope can be as large as $V\sim\pi r w^2/4$, such that $P_{\rm sub}\propto w^2$; thus it is proportional to the surface of the plasmoid. Using the angular size of the plasmoid as a proxy for the pulse width, $\delta\Phi\sim w/r$, yields $P_{\rm sub}\propto \delta\Phi^2$. Therefore, we obtain $F\propto \delta\Phi$, which is in qualitative agreement with the observed correlation. The bottom panel in Figure~\ref{fig:subpulse} reveals that the subpulse flux is distributed according to the power law $N_{\rm sub}\propto F^{-1}$  up to about a few times the standard deviation, beyond which the distribution cuts off. A similar distribution is observed for the subpulse  width, $N_{\rm sub}\propto \delta\Phi^{-1}$. This distribution is also consistent with the size distribution of plasmoids, $N\propto w^{-1}$, reported in Sect.~\ref{sect_plasmoids} \citep{uzd+10,sir+16}.

\section{Discussion and Conclusion} \label{sec:discandconc}

In this work, we make a case for the existence of bright subpulses in the incoherent synchrotron radiation emitted in the pulsar wind current layer (as opposed to the coherent radio emission). The origin of subpulses in the simulation is undeniably related to plasmoid formation in the layer. We can infer several clear falsifiable predictions from
this work. A first prediction is that the subpulse emission should appear mostly on the leading edge of each pulse. This phenomenon stems from the caustic effect in the current sheet. Although subpulses are drifting in phase, their motion is too fast to be observed over multiple pulsar periods by the same observer, in contrast to drifting subpulses observed in the radio band (e.g., \citealt{1968Natur.220..231D, 2007A&A...469..607W}), suggesting there is a different mechanism at work for the coherent emission (however, see \citealt{2019MNRAS.483.1731L, 2019ApJ...876L...6P}). New subpulses should therefore appear randomly after each period. The drift is due to the relativistic motion of plasmoids traveling away from the pulsar.

Another prediction is that the subpulse flux should be proportional to its width in phase: the brighter they are, the wider they should be in phase. A broad range of pulse flux is reported, but the variability is dominated by the largest plasmoids, which can sometimes be nearly as bright as the mean pulse profile. Although our limited numerical resources do not allow us to reproduce a realistic scale separation between the smallest and the biggest islands (e.g., meter-size to hundreds of kilometers in size in the Crab pulsar), our simulation probably captures the correct demographics of the largest plasmoids, and thus the largest subpulse variability reported here. Indeed, the hierarchical merging model of plasmoids combined with the effect of a relativistic expansion of the wind predicts that the number of large plasmoids is governed by the inverse of the dimensionless reconnection rate, $\beta^{-1}_{\rm rec}\gtrsim 10$ \citep{2021A&A...656A..91C}, which is consistent with the number of subpulses per pulse reported in this work. The duration of the brightest subpulses should then be $\sim\beta^{-1}_{\rm rec}$ times shorter than the main pulses. As the reconnection rate is nearly independent of plasma parameters in the ultra-relativistic regime ($\sigma\gg 1$, \citealt{2018MNRAS.473.4840W}), bright subpulses should be present in all rotation-powered pulsars. We also anticipate that our results will hold in full 3D where the overall picture of flux rope formation is qualitatively similar to 2D equatorial simulations \citep{2020A&A...642A.204C}. A possible caveat is that the contribution from other latitudes in 3D may reduce the level of predicted noise.

This work provides yet another observational test for the magnetic reconnection scenario and specifically for the origin of the pulsed incoherent emission. With the intrapulse variability being dominated by at most a dozen large subpulses, we speculate that these may be detectable. The predicted emission does not fall in the radio band, but instead at higher frequencies, from the infrared to the gamma-ray bands. At these frequencies, most extrinsic sources of noise such as propagation effects through the interstellar medium are absent, meaning that any intrinsic source of noise could in principle be more easily identified than in radio. However, a major limitation of this type of measurement will be the lack of photons and time resolution. Perhaps the most promising avenue would be optical observations of the Crab pulsar with a high time resolution \citep{ger+12}. X-ray observations by the NICER instruments may also provide important constraints on the intrinsic variability \citep{2021Sci...372..187E, 2022ApJ...928...67H}.

\begin{acknowledgements}
This work received funding and support from Campus France, the French Embassy in Ankara, the COST Action PHAROS (CA16214), and the European Research Council (ERC) under the European Union’s Horizon 2020 research and innovation programme (grant agreement No 863412). Simulations presented in this paper were performed using the GRICAD infrastructure (\url{https://gricad.univ-grenoble-alpes.fr}), which is supported by Grenoble research communities.
\end{acknowledgements}

\bibliography{refs.bib}
\bibliographystyle{aa}

\end{document}